\documentclass{ws-stone}
\def\etal{{\it et al.}}

\begin{document}

\title{Hadronic $B$ Decays to Charm from CLEO}

\author{Sheldon Stone}

\address{Physics Department, Syracuse University, Syracuse, N. Y. 13244-1130,
USA\\E-mail: stone@phy.syr.edu}

\twocolumn[\maketitle\abstract{Recent results include the measurement of
$B\to D^{(*)}\pi^+\pi^-\pi^-\pi^o$ decays, with rates of $\sim$1.8\%,
observation of a significant $\omega\pi^-$ substructure that appears to
result from the decay of the $\rho'$ resonance, whose mass and width we
determine as 1418 MeV and 382 MeV, respectively. Using $B$ to charmonium decays
we find that the $B^o$ fraction in $\Upsilon(4S)$ decays is nearly half, that the CP asymmetries in $J/\psi K^{\mp}$ and $\psi' K^{\mp}$ are 
small, and we also
report on the first observation of $B\to\eta_c K$.
}]

\section{Introduction }
Understanding hadronic decays of the $B$ is crucial to insuring
that decay modes used for measurement of CP violation truly reflect the
underlying quark decay mechanisms expected theoretically.
Most of the decay rate is hadronic. The semileptonic 
branching ratio for $e^-,~\mu^-$ and $\tau^-$ totals 
$\sim$25\%.\cite{PDG} 
Currently, measured exclusive branching ratios for hadronic $B$
decays total only a small fraction of the hadronic width. The measured
modes for the
$\overline{B}^o$, including $D^+ (n\pi)^-$, $D^{*+} (n\pi)^-$, where 
$3\ge n \ge 1$, $D^{+(*)}D_s^{-(*)}$ and $J/\psi$ exclusive, totals only 
about 10\%.\cite{PDG} Thus our understanding of hadronic $B$ decay modes is 
not yet well based in data.
It is also interesting to note that the average charged multiplicity in  
hadronic $B^o$ decays is 5.8$\pm 0.1$.\cite{chargedmult} Since this
multiplicity contains contributions from the $D^+$ or $D^{*+}$ normally 
present in $\overline{B}^o$ decay, we expect a sizeable, approximately several
percent, decay rate into final states with four pions.\cite{Argus} 

\section{
First Observation of the $\rho'$ in $B$ Decays}

We have made the first statistically significant observations of     
six hadronic $B$ decays shown in Table~\ref{table:brs} that result from
studying the reaction $B\to D^{(*)}\pi^+\pi^-\pi^-\pi^o$.\cite{ds4pi}
    
\begin{table} [htb]   
\begin{center}    
\caption{Measured Branching Ratios}    
\label{table:brs}    
\begin{tabular}{|l|c|}\hline    
\raisebox{0pt}[12pt][6pt]{Mode} & $\cal{B}$ (\%) \\\hline    
\raisebox{0pt}[12pt][6pt]{$\overline{B}^o\to D^{*+}\pi^+\pi^-\pi^-\pi^o$} &     
1.72$\pm$0.14$\pm$0.24    \\\hline    
\raisebox{0pt}[12pt][6pt]{$\overline{B}^o\to D^{*+}\omega\pi^-$} & 0.29$\pm$0.03$\pm$0.04     
\\\hline    
\raisebox{0pt}[12pt][6pt]{$\overline{B}^o\to D^{+}\omega\pi^-$} & 0.28$\pm$0.05$\pm$0.03       
  \\ \hline   
\raisebox{0pt}[12pt][6pt]{${B}^-\to D^{*o}\pi^+\pi^-\pi^-\pi^o$}&1.80$\pm$0.24$\pm$0.25           
   \\ \hline   
\raisebox{0pt}[12pt][6pt]{${B}^-\to D^{*o}\omega\pi^-$}  & 0.45$\pm$0.10$\pm$0.07                 
\\ \hline   
\raisebox{0pt}[12pt][6pt]{${B}^-\to D^{o}\omega\pi^-$}  &0.41$\pm$0.07$\pm$0.04        
     \\    
\hline    
\end{tabular}    
\end{center}    
\end{table}    
    
\begin{figure*}[t]    
\epsfxsize30pc
\figurebox{16pc}{32pc}{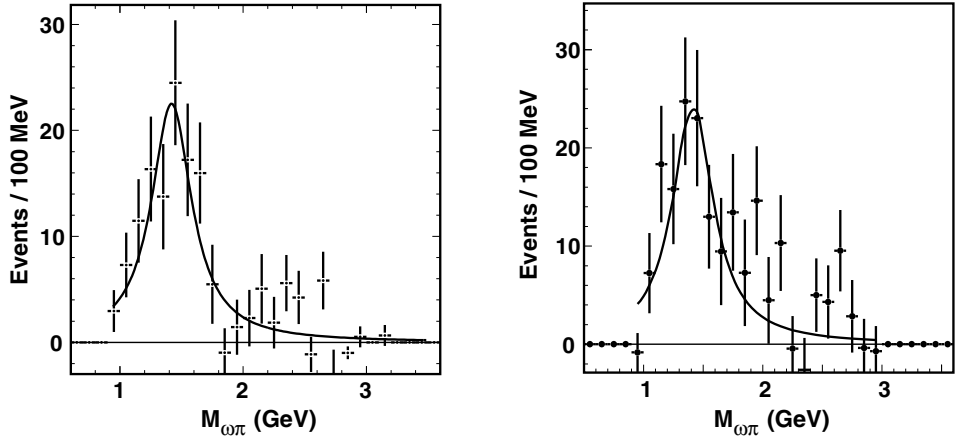}    
\caption{\label{omega_pi}(left) The background subtracted $\omega\pi^-$ mass spectrum   
from $\overline{B}^o\to D^{*+}\omega\pi^-$ decays fit to a simple   
Breit-Wigner shape. (right)  Same for $B\to D\omega\pi^-$ decays 
($D^o$ and $D^+$ are summed).}    
\end{figure*}

A significant fraction of the final state is $D^{(*)}\omega\pi^-$, and    
there is a low-mass resonant substructure in the $\omega\pi^-$     
mass. (See Fig.~\ref{omega_pi}) A simple Breit-Wigner fit assuming a single 
resonance and no     
background gives a mass of 1418$\pm$26$\pm$19 MeV with an intrinsic width     
of 382$\pm$41$\pm$32 MeV.    
 
We determine the spin and parity of the $\omega\pi^-$ resonance (denoted
$A$ temporarily) by considering the decay sequence $B\rightarrow A\ D$; 
$A\rightarrow
\omega \pi$ and $\omega \rightarrow \pi ^+ \pi ^- \pi ^o$.

\begin{figure*}[bth]    
\epsfxsize33pc
\figurebox{16pc}{32pc}{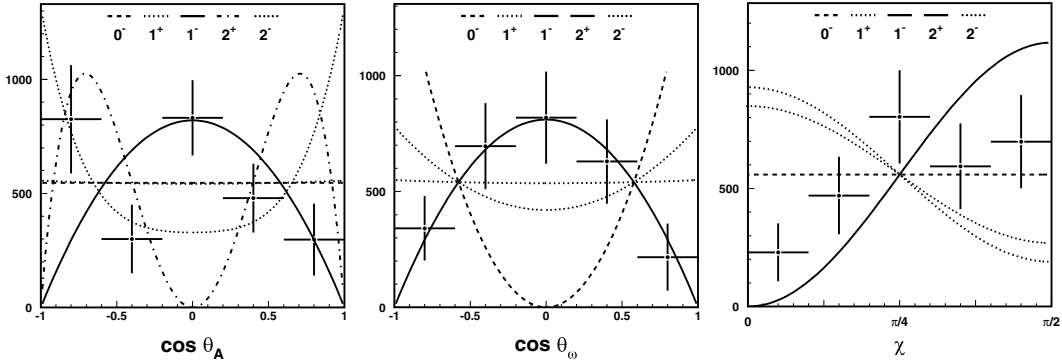}    
\vspace{.2cm}  
\caption{\label{helicity_angles}The angular distribution of 
$\theta_A$ (top-left),
$\theta_\omega$ (top-right) and $\chi$ (bottom).
 The curves show the best fits to the data for 
      for different $J^P$ assignments. The $0^-$ and $1^+$ are 
almost
      indistinguishable in $\cos\theta_A$, while the $1^-$ and 
$2^+$ are indistinguishable in $\cos\theta_{\omega}$ and $\chi$.
The vertical axis gives efficiency corrected events, 104 events are used.}    
\end{figure*}

The angular distributions are shown in Fig.~\ref{helicity_angles}. Here
$\theta_A$ is the angle between the $\omega$ direction in the $A$ rest frame
and the $A$ direction in the $B$ rest frame; 
$\theta _{\omega}$ is the orientation of 
the
$\omega$ decay plane in the $\omega$ rest frame, and $\chi$ is the angle
between the $A$ and $\omega$ decay planes.

The data are fit to the expectations for the various $J^P$ 
assignments. 
The $\omega$ polarization is very clearly 
transverse 
($\sin^2\theta_{\omega}$) and that infers a $1^-$ or $2^+$ 
assignment. The $\cos\theta_A$ distribution prefers $1^-$, as does
the fit to all three projections.  
Thus this state is likely to be the elusive $\rho'$     
resonance.\cite{Clegg}     
These are by far the most accurate and     
least model dependent measurements of the $\rho'$ parameters.    
The $\rho'$ dominates the final state. (Thus the branching ratios     
for the $D^{(*)}\omega\pi^-$ apply also for $D^{(*)}\rho'^-$.)

Heavy quark symmetry predicts equal partial widths for $D^*\rho'$ and
$D\rho'$. We measure the relative rates to be     
$1.06 \pm 0.17 \pm 0.04~;$       
consistent within our relatively large errors.     
    
\begin{figure}
\epsfxsize150pt
\figurebox{120pt}{160pt}{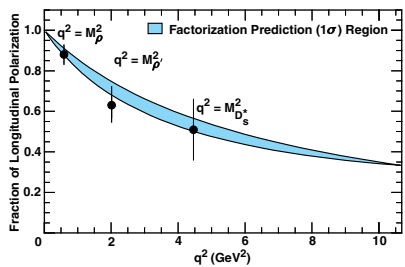}
\caption{Measured $D^{*+}$ polarization versus model predictions.}
\label{polar_bw}
\end{figure}

Factorization predicts that the fraction of longitudinal      
polarization of the $D^{*+}$ is the same as in the related     
semileptonic decay $B\to D^*\ell^-\bar{\nu}$ at four-momentum     
transfer $q^2$ equal to the mass-squared of the $\rho'$    
\begin{equation}    
{{\Gamma_L\left({B}\to D^{*+}\rho'^-\right)}\over     
{\Gamma\left({B}\to D^{*+}\rho'^-\right)}} =     
{{\Gamma_L\left({B}\to D^{*}\ell^-\bar{\nu}\right)}\over     
{\Gamma\left({B}\to D^{*}\ell^-    
\bar{\nu}\right)}}\left|_{q^2=m^2_{\rho'}}\right.~~.    
\end{equation}    
   Our measurement of the $D^{*+}$ polarization  is     
 (63$\pm$9)\%.     
The model predictions in semileptonic decays for a $q^2$ of 2     
GeV$^2$, are between    
66.9 and 72.6\%.\cite{slmodels} Fig.~\ref{polar_bw} shows the measured
polarizations for the $D^{*+}\rho'^-$, the $D^{*+}\rho^-$,\cite{rhopol} and the
$D^{*+}D_s^{*-}$ final states. The latter is based on a new measurement using
partial reconstruction of the $D^{*+}$.\cite{DD} 
Thus this prediction of factorization is     
satisfied.     

\section{New Results from $B\to Charmonium$ Decays}
\subsection{\boldmath ${f_{oo}\over f_{+-}}$=${{{\cal B}\left(\Upsilon(4S)\to B^o \overline{B}^o\right)}\over   
{{\cal B}\left(\Upsilon(4S)\to B^+B^-\right)}}$}   
    
We measured an important   
although often overlooked, parameter that effects every $B^o$ or $B^+$     
branching ratio measured on the $\Upsilon (4S)$ resonance. We determined     
the ratio of the number of $B^o\overline{B}^o$ pairs versus the     
number of $B^-B^+$ pairs by measuring the ratio of widths of two-body     
final states of the type $\psi^{(')}K^{(*)}$. Since the $\psi$ is I=0 and the     
$B$ and $K$ are I=1/2, the decay is pure $\Delta$I=1/2 and therefore the     
individual channels, such as $\psi K$, must have the same decay widths for     
$B^o$ and $B^+$. We find ${f_{oo}\over f_{+-}}$= 1.04$\pm$0.07$\pm$0.04,   
which yields nearly equal rates for charged and neutral $B$'s, i.e.    
$f_{oo}=0.49 \pm 0.02 \pm 0.01$~.\cite{Silvia}

\subsection{Limits on CP Violation in 
{\boldmath $B^{\mp}\to J/\psi$ (or $\psi'$) $K^{\mp}$}}

In the Standard Model only one decay amplitude dominates for decays of the type
$B^{\mp}\to$ {\it charmonium} $K^{\mp}$. CLEO measured asymmetries consistent
with zero for the $J/\psi$ and $\psi'$ cases: (1.8$\pm$4.3$\pm$
0.4)\% and (2.0$\pm$9.1$\pm$1.0)\%, respectively.\cite{CPA}

\subsection{First Measurement of ${\cal{B}}(B\to \eta_c K$)}

We observe exclusive final states with the $c\overline{c}$ being
the $\eta_c$.\cite{etac} The rates are ${\cal{B}}(B^-\to \eta_c K^-)=
(6.9^{+2.6}_{-2.1}\pm 0.8\pm 2.0)\times 10^{-3}$ and ${\cal{B}}(B^o\to \eta_c K^o)=
(10.9^{+5.5}_{-4.2}\pm 1.2\pm 3.1)\times 10^{-3}$. These rates are large, similar to those
for the $J/\psi K$ final state.

\section{Conclusions}
Recent CLEO data has vastly increased our knowledge about hadronic $B$ decays.
We have seen the elusive $\rho'$ and have the best
measurement of its mass and width. Factorization when applied to polarizations
appears to work in $D^{*+}$ plus $\rho$, $\rho'$ and $D_s^*$ decays.
The rates of the $\Upsilon(4S)$ into charged and neutral $B$'s has been
accurately determined and the first observation of $B\to \eta_c K$ decays has
been made. Much however remains to be learned.

\section*{Acknowledgments}
We thank N. Isgur, J. Rosner and J. Schechter for useful discussions.
We thank the National Science Foundation for support.

\end{document}